\documentclass[a4paper,11pt]{article}
\usepackage{epsfig,amssymb,amsfonts,amsmath,mathtools,bm,color}
\usepackage{jheppub}
\newcommand{\be}{\begin{equation}}
\newcommand{\ee}{\end{equation}}
\newcommand{\bea}{\begin{eqnarray}}
\newcommand{\eea}{\end{eqnarray}}
\newcommand{\beas}{\begin{eqnarray*}}
\newcommand{\eeas}{\end{eqnarray*}}
\newcommand{\ds}{\displaystyle}
\newcommand{\ven}{{\bm n}}
\newcommand{\vep}{{\bm p}}

\newcommand{\venu}{{\bm \nu}}
\newcommand{\Lag}{\mathcal{L}}
\newcommand{\Tr}{\textrm{Tr}}

\newcommand{\Hb}{\bar{H}}

\newcommand{\veV}{{\bm V}}
\newcommand{\vesig}{{\bm \sigma}}

\graphicspath{{Figs.dir/}}
\title{\boldmath Can $X(3915)$ be the tensor partner of the $X(3872)$?}

\author[a,b]{V. Baru}
\author[c]{C. Hanhart}
\author[b,d,e]{A. V. Nefediev}

\affiliation[a]{Institut f\"ur Theoretische Physik II, Ruhr-Universit\"at Bochum, D-44780 Bochum, Germany}
\affiliation[b]{Institute for Theoretical and Experimental Physics, B. Cheremushkinskaya 25, 117218 Moscow, Russia}
\affiliation[c]{Forschungszentrum J\"ulich, Institute for Advanced Simulation, Institut f\"ur Kernphysik and
J\"ulich Center for Hadron Physics, D-52425 J\"ulich, Germany}
\affiliation[d]{National Research Nuclear University MEPhI, 115409, Kashirskoe highway 31, Moscow, Russia}
\affiliation[e]{Moscow Institute of Physics and Technology, 141700, Institutsky lane 9, Dolgoprudny, Moscow Region, Russia}

\emailAdd{vadimb@tp2.rub.de}
\emailAdd{c.hanhart@fz-juelich.de}
\emailAdd{nefediev@itep.ru}

\abstract{
It has been proposed recently (Phys.\ Rev.\ Lett.\ {\bf 115} (2015), 022001) that the charmoniumlike state named $X(3915)$ and suggested to be a $0^{++}$ scalar, 
is just
the helicity-0 realisation of the $2^{++}$ tensor state $\chi_{c2}(3930)$. This scenario would call for a helicity-0 dominance, which were at odds with the 
properties of a conventional tensor charmonium, but might be compatible with some exotic structure of the $\chi_{c2}(3930)$.
In this paper, we investigate, if such a scenario is compatible with the assumption that the $\chi_{c2}(3930)$ is a 
$D^*\bar D^*$ molecular state --- a spin partner of the $X(3872)$ treated as a shallow bound state. 
We demonstrate that for a tensor molecule the helicity-0 component vanishes for vanishing binding energy
and accordingly for a shallow bound state a helicity-2 dominance would be natural.
However, for the $\chi_{c2}(3930)$, residing about $100$ MeV below the $D^*\bar D^*$ threshold, there is no a priori reason for a helicity-2 dominance 
and thus the proposal formulated in the above mentioned reference might indeed point at a molecular
structure of the tensor state. Nevertheless, we find that the experimental data currently available
favour a dominant contribution of the helicity-2 amplitude
also in this scenario, if spin symmetry arguments are employed to relate
properties of the molecular state to those of the $X(3872)$. 
We also discuss what research is necessary to further constrain the analysis.
}

\begin{document} 
\maketitle
\flushbottom


\section{Introduction}

Among the most interesting and intriguing discoveries made in high energy physics in recent years, one should mention the discovery of many 
new hadrons lying above the open-flavour threshold both in the spectrum of charmonium and bottomonium --- for reviews see, for example,
Refs.~\cite{Brambilla:2010cs,Brambilla:2014jmp,Abe:2010gxa,Drutskoy:2012gt,Asner:2008nq,Lutz:2009ff}. One of such states, the $X(3915)$, was observed by 
the Belle Collaboration in the two-photon annihilation to the $\omega J/\psi$ final state~\cite{Uehara:2009tx}, and the variety of options for the 
$J^{PC}$ quantum numbers of this state was limited to just $0^{++}$ and $2^{++}$. Later, the BaBar Collaboration reported that the angular distributions 
for the final-state leptons and pions emerging from the decays of the $J/\psi$ and $\omega$ favoured the $0^{++}$ option~\cite{Lees:2012xs}, so that 
this state is conventionally identified as the $\chi_{c0}(2P)$ charmonium \cite{Liu:2009fe,Olive:2016xmw} --- although
this charmonium assignment was questioned in Refs.~\cite{Guo:2012tv,Olsen:2014maa}. 
In Ref.~\cite{Li:2015iga} the $X(3915)$ was proposed to be 
a scalar $D_s\bar{D}_s$ molecule. 

However, it was noticed recently
\cite{Zhou:2015uva} that the assumption used by BaBar in the data analysis, namely the assumption of a helicity-2 dominance 
for the tensor state, might not hold if the tensor state had an exotic structure. Acknowledging this, the state is called $X(3915)$ in the
2016 Review of Particle Physics (RPP) by the PDG~\cite{Olive:2016xmw}. 

Historically, it was found long ago \cite{Alekseev} that, in the two-photon decays 
of the $2^{++}$ positronium, 
only the helicity-2 amplitude contributes while the helicity-0 amplitude vanishes. This observation was later generalised to quarkonia 
\cite{Krammer:1977an,Li:1990sx} ,
and the helicity-0 amplitude was demonstrated to provide only a small relativistic correction to the dominating helicity-2 amplitude,
in agreement with the findings of Ref.~\cite{Alekseev}. However, a similar analysis for 
exotic structures has not been done so far. Thus, it was pointed out in 
Ref.~\cite{Zhou:2015uva} that the helicity-2 dominance constraint may be relaxed in 
the data analysis if one assumes the $X(3915)$ to be some exotic state. The authors concluded that, for the helicity-0 amplitude 
comparable in magnitude with the helicity-2 one, the measured angular distributions could be reproduced under the 
assumption of $2^{++}$ 
quantum numbers of the $X(3915)$ suggesting that what was observed in Ref.~\cite{Uehara:2009tx} was simply the
manifestation of the helicity-0 component of the tensor state known as $\chi_{c2}(3930)$.\footnote{Note that here and in what follows 
calling the state $\chi_{c2}(3930)$ only means the definition of its quantum numbers and it
does not imply a $c\bar{c}$ nature of this state, fully in line with the naming scheme defined 
in the RPP~\cite{Olive:2016xmw}.} 
In this paper, we investigate if a prominent helicity-0 component is compatible with $X(3915)$ being
a $2^{++}$ $D^*\bar D^*$ molecular state.
To this end, we briefly repeat the theoretical arguments why one should expect a $2^{++}$ $D^*\bar{D}^*$ 
molecule to exist as a spin partner of the $X(3872)$. Then we employ the Occam's razor principle to identify 
the $X(3915)$ with this hypothetical spin-2
partner --- the assumption allowing one to relate the effective coupling constant of the $S$-wave transition 
$X(3915)\to D^*\bar{D}^*$ to the experimentally measured 
binding energy of the $X(3872)$. Equipped with this information, we study the properties of the $X(3915)$ in the two-photon annihilation processes. 

One of the celebrated theoretical tools used in studies of hadronic states with heavy quarks is the Heavy-Quark Spin Symmetry (HQSS) which is based 
on the observation that, for $\Lambda_{\rm QCD}/m_Q\to 0$, with $m_Q$ denoting the quark mass, the strong interactions in the system are independent of 
the heavy quark spin. As a result, the hypothesis of the existence of a molecular state at one open-flavour threshold entails the existence of spin
partner states at the neighbouring open-flavour thresholds, which differ by the heavy-quark spin orientation. For example, the idea of the existence of spin 
partners for the isovector bottomonium-like states $Z_b^+(10610)$ and $Z_b^+(10650)$ was put forward and investigated in 
Refs.~\cite{Bondar:2011ev,Voloshin:2011qa,Mehen:2011yh}. Although in case of the charm quark the ratio $\Lambda_{\rm QCD}/m_c\simeq 0.2$ is 
sizable and one expects non-negligible corrections to the strict symmetry limit, constraints from HQSS can still provide a valuable guidance also in 
the charm sector and, in particular, for the $X(3872)$ \cite{Hidalgo-Duque:2013pva}. Thus, it was argued in Refs.~\cite{Nieves:2012tt,Guo:2013sya} that 
one should expect a shallow $S$-wave bound state in the $D^*\bar D^*$ channel with the quantum numbers $J^{PC}=2^{++}$ --- the molecular partner of the 
$X(3872)$ conventionally denoted as $X_2$. In Ref.~\cite{Albaladejo:2015dsa}, on the basis of an effective field theory with perturbative pions (X-EFT), 
the width of this 
state was 
estimated to be as small as a few MeV. Later, in Ref.~\cite{Baru:2016iwj}, an alternative EFT approach to the $X_2$ state was formulated 
considering pion exchanges nonperturbatively and it was concluded that the mass of this state might acquire a significant shift and that its width 
could be as large as several tens of MeV.
In particular, its binding energy was found to constitute a few dozens MeV. An exploratory study of the possible impact of the 
genuine quarkonium on the formation of this molecular state is presented in Ref.~\cite{Cincioglu:2016fkm} and a further shift of 
the corresponding pole into the complex plane was argued to be possible.

Therefore, although the measured mass of the $X(3915)$ lies approximately 100~MeV below the $D^*\bar{D}^*$ 
threshold \cite{Olive:2016xmw}, in the current research, we dare identify it with the $X_2$ --- 
the tensor spin 
partner of the $X(3872)$. Then, using the measured properties of the $X(3872)$ as an anchor, we 
trace the consequences of such an identification for the relative strength 
of the helicity-0 and helicity-2
amplitudes in the two-photon fusion processes proceeding through the formation of the $X(3915)$ state.

The paper is organised as follows: In Sec.~\ref{Sec:ggX2}
the amplitude for $\gamma\gamma\to X_2$ is decomposed into a complete set of four gauge invariant, mutually orthogonal tensor structures. 
We demonstrate how the helicity-0 and helicity-2 amplitudes can be expressed 
in terms of these tensors. 
In addition, the angular distributions of different two-photon annihilation processes proceeding via the $X_2$ state are evaluated and expressed 
in terms of the helicity amplitudes. We demonstrate that the leading helicity-0 amplitude 
contribution to the observables depends only on the $X_2\to D^*\bar{D}^*$ coupling
that may be estimated from the $X(3872)\to D\bar{D}^*$ coupling, as explained in Subsec.~\ref{sec:X2D*D*}, 
while the helicity-2 amplitude requires an additional contact term for renormalisation. 
Given that this contact term is unknown, the relative importance of the different helicity contributions for a tensor molecule 
cannot be fixed unambiguously without an additional experimental input --- this insight is new to the
best of our knowledge.

In order to proceed, we provide a complete evaluation of the analytic expressions 
derived for the helicity-0 amplitude (quoted in Appendix~\ref{app:hel0}) and 
involve additional plausible assumptions to extract the ratio of the helicity-2 to helicity-0 
amplitudes directly from experimental data (for completeness, we also quote the explicit form of the helicity-2 amplitude 
in Appendix~\ref{app:hel2}). To this end, in 
Subsec.~\ref{sec:X2D*D*}, we employ HQSS and the molecular 
interpretation for the $X(3872)$ and its spin partners 
to express the coupling constants of the spin-2 and spin-0 states to the $D^{(*)}\bar{D}^{(*)}$ meson pairs through the binding energy of 
the $X(3872)$ and, in Subsec.~\ref{Sec:GX2gg}, we confront the evaluated helicity-0 contribution to the $X_2$
two-photon decay width with the experimental data to draw conclusions on the relative importance of the helicity-0 and helicity-2 amplitudes. 
Our results point towards a helicity-2 dominance for the spin-2 heavy-quark spin symmetry partner of the
$X(3872)$, although, as stressed in Sec.~\ref{sec:DDfi}, due to limited information currently available about
this state, this conclusion is subject to potentially large uncertainties. Such a helicity-2 dominance is at odds with the need expressed in Ref.~\cite{Zhou:2015uva} 
that, in order to be consistent with a $2^{++}$ state, the angular distributions of the various two-photon annihilation processes call for a sizable 
helicity-0 contribution. Therefore, our study suggests that if the $X(3915)$ were indeed 
just a realisation of the $2^{++}$ state $\chi_{c2}(3930)$, its properties seem to be not consistent with its being 
the predicted spin partner of the $X(3872)$.
On the other hand, if the $X(3915)$ is a molecular partner of the $X(3872)$, it is presumably the 
scalar state. We also discuss how additional data would allow one to draw more firm conclusions.

\section{The amplitude $\gamma\gamma\to X_2$}
\label{Sec:ggX2}

\subsection{Electric and magnetic contributions}

\begin{figure}[t]
\begin{center}
\epsfig{file=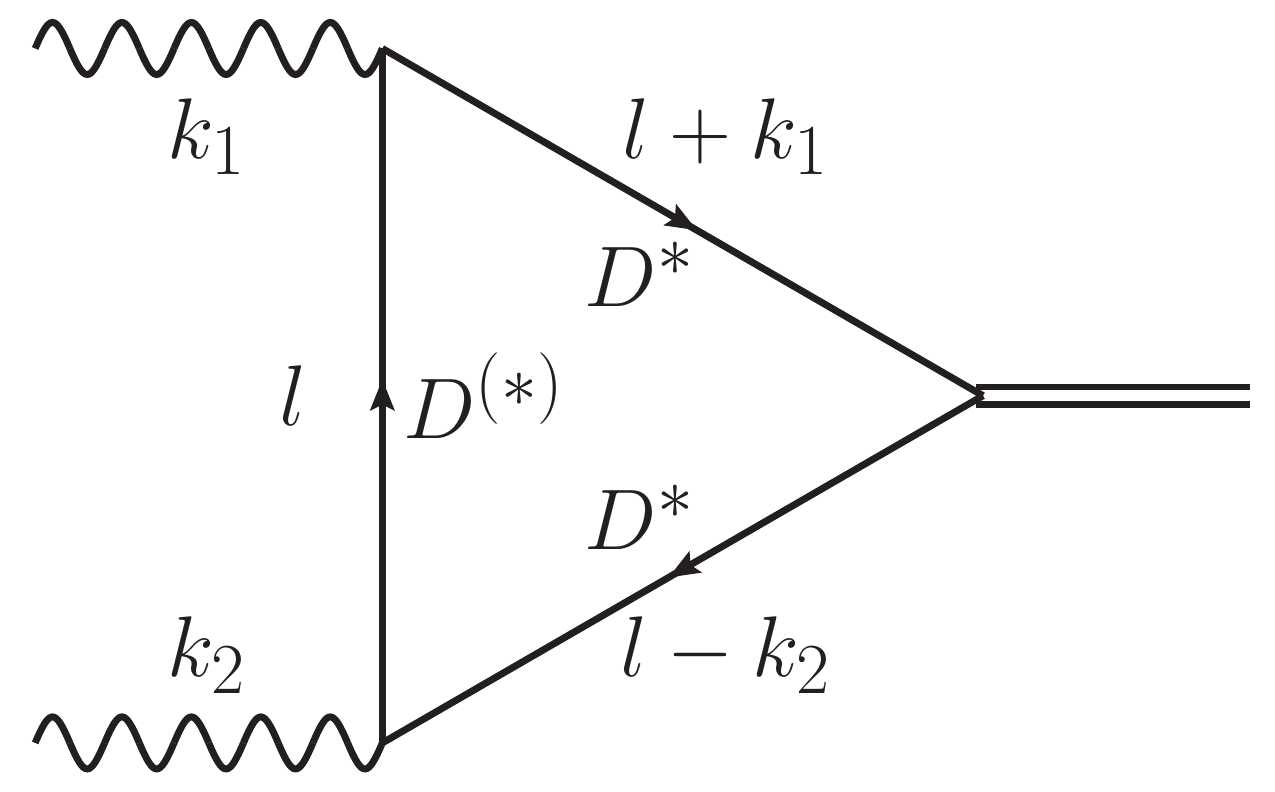,width=0.3\textwidth}\hspace*{5mm}
\epsfig{file=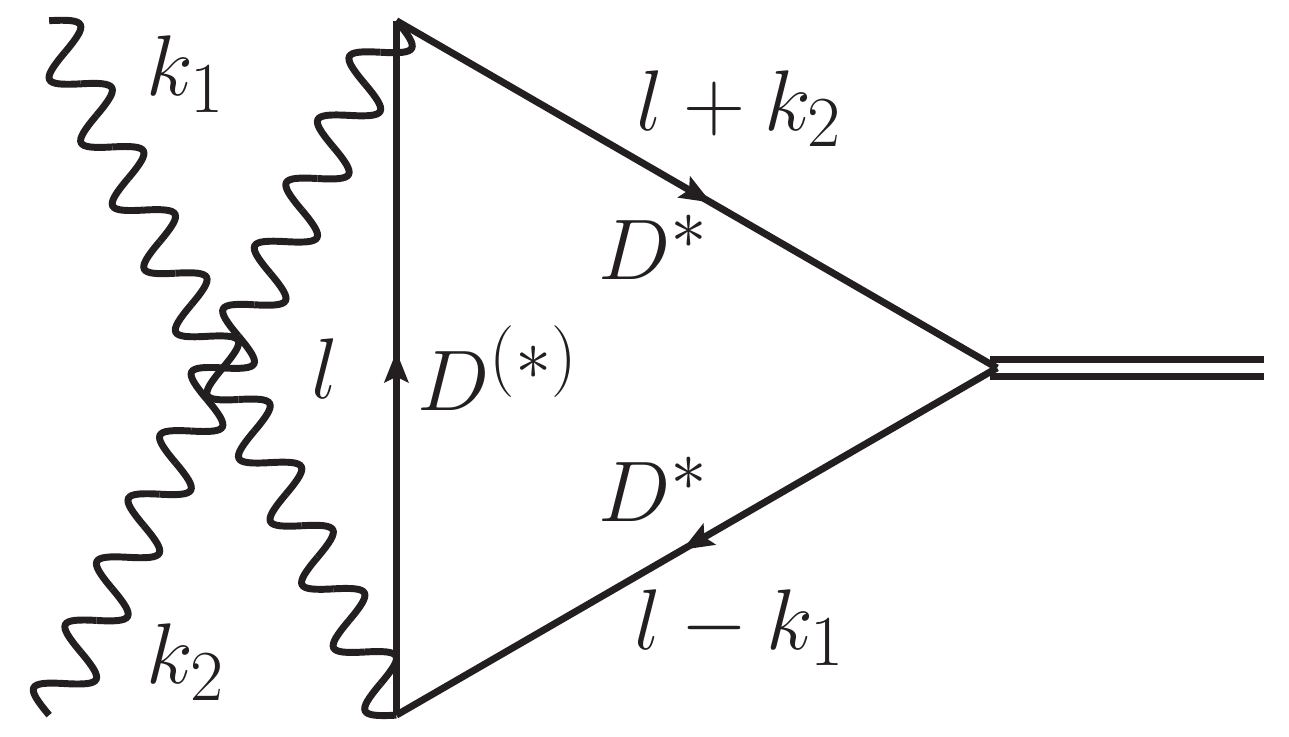,width=0.3\textwidth}\hspace*{5mm}
\epsfig{file=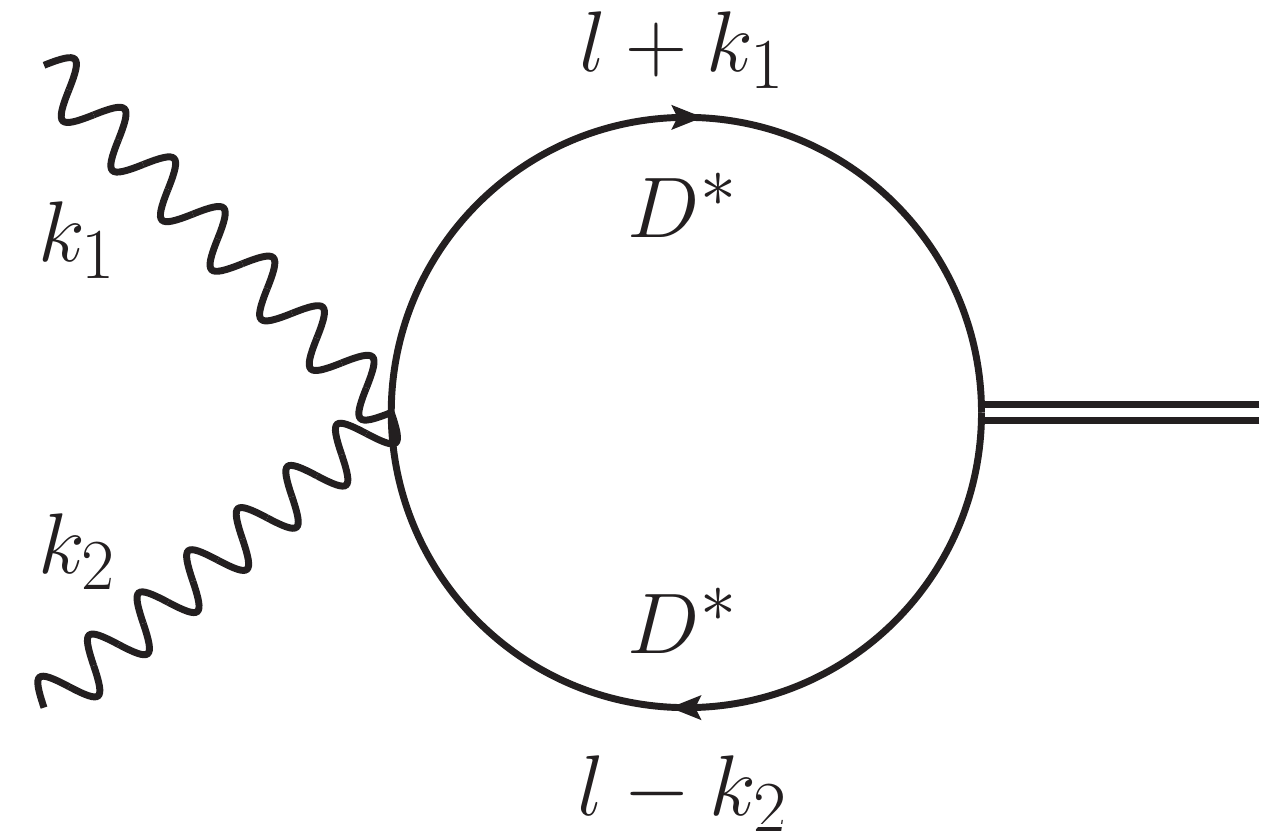,width=0.27\textwidth}
\end{center}
\caption{Diagrams contributing to the amplitude (\ref{amplitude1}). }\label{fig:diagramsmagn}
\end{figure}

The amplitude of the fusion process $\gamma\gamma\to X_2$ can be written in the form
\be
M(\gamma\gamma\to X_2)=M^{\mu\nu\rho\sigma}\varepsilon_\mu(k_1)\varepsilon_\nu(k_2)\varepsilon_{\rho\sigma}(p),\quad p=k_1+k_2,
\label{amplitude1}
\ee
where $\varepsilon_\mu(k_1)$, $\varepsilon_\nu(k_2)$ are the first and the second photon polarisation vector, respectively, and 
$\varepsilon_{\rho\sigma}$ is the $X_2$ polarisation tensor which obey the standard constraints,
\be
k_1\cdot\varepsilon(k_1)=k_2\cdot\varepsilon(k_2)=0,\quad p^\rho\varepsilon_{\rho\sigma}(p)=p^\sigma\varepsilon_{\rho\sigma}(p)=
g^{\rho\sigma}\varepsilon_{\rho\sigma}(p)=0.
\label{tensprop}
\ee

There are two mechanisms responsible for the $D^{(*)}$ interaction with the electromagnetic field which produce two types of 
vertices: the electric vertices and the magnetic ones. 
The covariant form of the electric $D^{*a}_{\mu}(p_1) \to D^{*b}_{\nu}(p_2)\gamma_{\lambda}(q)$ ($p_1=p_2+q$) vertex reads \cite{Guo:2014taa}
\be
\Gamma^{(e)ab}_{\mu\nu\lambda}(p_1,p_2)=\langle D^{*b}_{\nu}(p_2)\gamma_{\lambda}(q)|\Lag_e|D^{*a}_{\mu}(p_1)\rangle
=e\Bigl[(p_1+p_2)_{\lambda}g_{\mu\nu}-p_{1\nu}g_{\mu\lambda}-p_{2\mu}g_{\nu\lambda}\Bigr] Q_D^{ab},
\label{psipsigam}
\ee
where $\Lag_e$ is the electric part of the interaction Lagrangian and $\hat{Q}_D=\text{diag}(0,1)$ is the charge matrix 
corresponding to the isospin doublet $D^{(*)}=(D^{(*)0},D^{(*)+})$; $\Gamma^{(e)ab}_{\mu\nu\lambda}(p_1,p_2)$ satisfies the Ward identity, 
\be
q^\lambda\Gamma^{(e)ab}_{\mu\nu\lambda}(p_1,p_2)=e\Bigl[(S^{-1}(p_2))_{\mu\nu}-(S^{-1}(p_1))_{\mu\nu}\Bigr]Q_D^{ab},
\nonumber
\ee
where the $D^*$ propagator and its inverse form are
\be
S_{\mu\nu}(p)=\frac{1}{p^2-m_*^2+i\varepsilon}\left(-g_{\mu\nu}+\frac{p_{\mu}p_{\nu}}{m_*^2}\right),
\quad
(S^{-1}(p))_{\mu\nu}=-(p^2-m_*^2)g_{\mu\nu}+p_{\mu}p_{\nu}.
\label{SS}
\ee

An additional electric seagull-like contact vertex $\gamma_\mu(k_1)\gamma_\nu(k_2)\to D_\alpha^{*a}(p_1)\bar{D}_\beta^{*b}(p_2)$ reads
\be
\Gamma^{(e)ab}_{\mu\nu\alpha\beta}
=\langle D_\alpha^{*a}(p_1)\bar{D}_\beta^{*b}(p_2)|\Lag_e|\gamma_\mu(k_1)\gamma_\nu(k_2)\rangle
=e^2\left(g_{\mu\alpha}g_{\nu\beta}+g_{\mu\beta}g_{\nu\alpha}-2g_{\mu\nu}g_{\alpha\beta}\right) Q_D^{ab}.
\label{fishe}
\ee

The Lagrangian describing the leading magnetic interaction between the $D$ and $D^*$ mesons takes the form
\be
\Lag_m=iem_*F_{\mu\nu} D_a^{*\mu\,\dag}{\cal M}_{D^*D^*}^{ab}D^{*\nu}_b 
+e\sqrt{mm_*}\epsilon_{\lambda\mu\alpha\beta}v^\alpha\partial^\beta
A^\lambda\left[D_a^{*\mu\,\dag}{\cal M}_{D^*D}^{ab}D_b+\mbox{h.c.}\right],
\label{Lagm}
\ee
where $v^\mu$ is the four-velocity of the heavy quark ($v^\mu v_\mu =1$) and the magnetic moment matrices for the $D^*\to D^{(*)}\gamma$ 
transitions read
\be
\hat{\cal M}_{D^*D^*}=\beta \hat{Q}-\frac{Q_c}{m_c}\hat{1},\quad
\hat{\cal M}_{D^*D}=\beta \hat{Q}+\frac{Q_c}{m_c}\hat{1},
\label{Mmatr}
\ee
with $Q=\text{diag}(2/3,-1/3)$ being the light-quark charge matrix, and $m_c$ and $Q_c$ being the charmed-quark mass and charge, $Q_c=2/3$, 
respectively. Here the leading $\beta$-terms account for the nonperturbative light-flavour cloud in the charmed meson and
the subleading terms proportional to $Q_c/m_c$ come from the magnetic moment
of the charm quark. In what follows we use the values~\cite{Hu:2005gf}
\be
\beta^{-1}=379~\mbox{MeV},\quad m_c=1863~\mbox{MeV}.
\ee

The nonrelativistic reduction of Eq.~(\ref{Lagm}) yields the Lagrangian 
which agrees with that given in Ref.~\cite{Hu:2005gf}.

Then, the magnetic $D^{*a}_{\mu}(k_1) \to D^{*b}_{\nu}(k_2)\gamma_{\lambda}(q)$
and $D^{*a}_{\mu}(k_1) \to D^b(k_2)\gamma_{\lambda}(q)$ ($k_1=k_2+q$) vertices read \cite{Guo:2014taa}
\be
\Gamma_{\mu\nu\lambda}^{(m)ab}(q)=\langle D^{*b}_{\nu}(k_2)\gamma_{\lambda}(q)|\Lag_m|D^{*a}_{\mu}(k_1)\rangle
=-em_*(q_\mu g_{\nu\lambda}-q_\nu g_{\mu\lambda}){\cal M}_{D^*D^*}^{ab}
\label{DsDsgamm}
\ee
and
\be
\Gamma_{\mu\lambda}^{(m)ab}(q)
=\langle D^b(k_2)\gamma_{\lambda}(q)|\Lag_m|D^{*a}_{\mu}(k_1)\rangle
=-ie\sqrt{mm_*}\varepsilon_{\mu\lambda\alpha\beta}v^{\alpha}q^\beta {\cal M}_{D^*D}^{ab},
\label{DsDgamm}
\ee
respectively. Both vertices are manifestly transversal with respect to the photon momentum $q^\lambda$.

The diagrams contributing to the $\gamma\gamma\to X_2$ fusion amplitude are depicted in Fig.~\ref{fig:diagramsmagn}. 
The first two diagrams acquire contributions from both electric and magnetic vertices given in Eq.~(\ref{psipsigam}) 
and in Eqs.~(\ref{DsDsgamm}), (\ref{DsDgamm}), respectively, while the last, fish-like diagram 
is purely electric --- see the vertex given in Eq.~(\ref{fishe}). It has to be noticed that the dominating contribution to this amplitude comes from the 
triangle diagrams with two magnetic vertices. Indeed, the magnetic photon emission vertices $D^*\to D^{(*)}\gamma$ contain the matrices (\ref{Mmatr})
which provide large numeric enhancement factors due to the isospin traces along the loop with two such vertices,
\be
\bar{m}^2\Tr\hat{\cal M}_{D^*D}^2\approx 18,\quad \bar{m}^2\Tr\hat{\cal M}_{D^*D^*}^2\approx 14,
\ee
which are to be confronted with the factors
\be
\bar{m}\Tr\left(\hat{\cal M}_{D^*D^*}\hat{Q}_D\right)\approx -2.4,\quad \Tr(\hat{Q}_D)=1,
\ee
for the triangle diagrams with one and with no magnetic vertex, respectively. Here the averaged mass $\bar{m}=(3m_*+m)/4\approx 1974$~MeV was used for the estimate.
In the evaluation above it was used that the $X(3915)$ is an isosinglet state. 

\subsection{Helicity decomposition}

There are in total four independent, gauge invariant two-photon tensors, which one may choose as
\bea
&&\hspace*{-10mm}{\cal S}_{\rho\sigma}^{(1)}=
e^2 h_1(s) g_{\rho\sigma} (\partial_\alpha F^{(1)}_{\mu\nu}) (\partial^\alpha F^{(2)\mu\nu}), \label{S1}\\
&&\hspace*{-10mm}{\cal S}_{\rho\sigma}^{(2)}=e^2 h_2(s) \left[(\partial_\rho F^{(1)}_{\mu\nu})(\partial_\sigma F^{(2)\mu\nu})
{+}(\partial_\sigma F^{(1)}_{\mu\nu})(\partial_\rho F^{(2)\mu\nu})
{-}\ds\frac12g_{\rho\sigma}(\partial_\alpha F^{(1)}_{\mu\nu})(\partial^\alpha F^{(2)\mu\nu})\right], \label{S2}\\
&&\hspace*{-10mm}{\cal S}_{\rho\sigma}^{(3)}=e^2 h_3(s) \left[(\partial_\rho\partial_\sigma F^{(1)}_{\mu\nu})F^{(2)\mu\nu}
+F^{(1)}_{\mu\nu}(\partial_\rho\partial_\sigma F^{(2)\mu\nu})\right], \label{S3}\\
&&\hspace*{-10mm}{\cal S}_{\rho\sigma}^{(4)}=e^2 h_4(s)\left[ F^{(1)}_{\rho\beta} F^{(2)\beta}_\sigma
+ F^{(1)}_{\sigma\beta} F^{(2)\beta}_\rho -\ds\frac12g_{\rho\sigma} F^{(1)}_{\mu\nu} F^{(2)\mu\nu} \right]\label{S4} .
\eea
The various terms in the interaction Lagrangian for the spin-2 field $X_2$ coupled to two photons are expressed as contractions 
of the above structures with the tensor $X_2^{\rho\sigma}$. Furthermore, the $h_n(s)$ ($n=\overline{1,4}$) in Eqs.~(\ref{S1})-(\ref{S4}) stand for scalar
functions that parametrise the dynamics of the decay.\footnote{Hereinafter we stick to the on-shell photons, so that 
($k_1^2=k_2^2=0$, $(k_1\cdot k_2)=s/2$).}
The set (\ref{S1})-(\ref{S4}) exhausts all possible second-order tensor 
structures built with the help of the two photon field tensors and symmetric with respect to the $\sigma\leftrightarrow\rho$ permutation. This 
guarantees that the set of the structures (\ref{S1})-(\ref{S4}) is complete in the given class of tensors.

In order to perform the helicity decomposition of the amplitude (\ref{amplitude1}) 
we introduce a quartet of tensors based on the structures from Eqs.~(\ref{S1})-(\ref{S4}),
\bea
&&\hspace*{-5mm}e_1^{\mu\nu\rho\sigma}=\frac{1}{2\sqrt{2}}g^{\rho \sigma }\left(g^{\mu\nu}-\frac{k_1^\nu k_2^\mu}{(k_1\cdot k_2)}\right),\label{e1}\\
&&\hspace*{-5mm}e_2^{\mu\nu\rho\sigma}=\frac{1}{\sqrt{2}}\left(\frac{k_1^\rho k_2^\sigma+k_1^\sigma k_2^\rho}{(k_1\cdot k_2)}-\frac12 g^{\rho\sigma}\right)
\left(g^{\mu\nu}-\frac{k_1^\nu k_2^\mu}{(k_1\cdot k_2)}\right),\label{e2}\\
&&\hspace*{-5mm}e_3^{\mu\nu\rho\sigma}=\frac{\left(k_1^\rho k_1^\sigma+k_2^\rho k_2^\sigma\right)}{2(k_1\cdot k_2)}
\left(g^{\mu\nu}-\frac{k_1^\nu k_2^\mu}{(k_1\cdot k_2)}\right),\label{e3}\\
&&\hspace*{-5mm}e_4^{\mu\nu\rho\sigma}=\frac{1}{2\sqrt{2}}\left(
\frac{k_{1\sigma}k_{2\mu}g_{\nu\rho}+k_{1\rho}k_{2\mu}g_{\nu\sigma}-(k_{1\sigma}k_{2\rho}+k_{1\rho}k_{2\sigma})g_{\mu\nu}}{(k_1\cdot k_2)}
\right.\nonumber\\
&&-\left.g_{\mu\sigma}\left(g^{\nu\rho}-\frac{k_1^\nu k_2^\rho}{(k_1\cdot k_2)}\right)
-g_{\mu\rho}\left(g^{\nu\sigma}-\frac{k_1^\nu k_2^\sigma}{(k_1\cdot k_2)}\right)
+g_{\rho\sigma}\left(g^{\mu\nu}-\frac{k_1^\nu k_2^\mu}{(k_1\cdot k_2)}\right)
\right),\label{e4}
\eea
which are mutually orthogonal and normalised,
\be
e_m^{\mu\nu\rho\sigma}e_n^{\mu\nu\rho\sigma}=\delta_{mn},\quad m,n=\overline{1,4}.
\ee
In addition, they are symmetric and transversal,
\be
e_n^{\mu\nu\rho\sigma}=e_n^{\mu\nu\sigma\rho},\quad k_{1\mu}e_n^{\mu\nu\rho\sigma}=k_{2\nu}e_n^{\mu\nu\rho\sigma}=0,\quad n=\overline{1,4}.
\label{props}
\ee

The set (\ref{e1})-(\ref{e4}) is complete,
so that the tensor $M^{\mu\nu\rho\sigma}$ defined in Eq.~(\ref{amplitude1}) can be decomposed as
\be
M^{\mu\nu\rho\sigma}=g_{X_2D^*\bar{D}^*}e^2\sum_{n=1}^4 C_n e_n^{\mu\nu\rho\sigma},
\label{Cndef}
\ee
where for later convenience the couplings are pulled out of the definition of the coefficients $\{C_n\}$.
 Gauge invariance and the $\rho\leftrightarrow\sigma$ 
symmetry of the tensor amplitude (\ref{Cndef}) are obvious due to the properties listed in Eq.~(\ref{props}). 
Therefore, the entire information about the amplitude (\ref{amplitude1}) is encoded in the functions $C_n$.

\section{Two-photon annihilation through the $X_2$}
\label{Sec:ggannihil}

Consider the two-photon annihilation process into various final states which proceeds through the formation of the $X_2$.
 Then, the tensor $M^{\mu\nu\rho\sigma}$ from Eq.~(\ref{amplitude1}) is to be contracted with the $X_2$ propagator taken in the standard form,
\be
G_{\rho\sigma,\alpha\beta}=\frac{{\cal P}_{\rho\sigma,\alpha\beta}}{p^2-M_{X_2}^2+iM_{X_2}\Gamma},
\ee
where 
\be
{\cal P}_{\rho\sigma,\alpha\beta}=\sum_{\xi=-2}^2\varepsilon_{\rho\sigma}^{(\xi)}(p)\varepsilon_{\alpha\beta}^{(\xi)}(p)=
\frac12\left(P_{\rho\alpha}P_{\sigma\beta}+P_{\rho\beta}P_{\sigma\alpha}
-\frac23P_{\rho\sigma}P_{\alpha\beta}\right),
\ee
$$
P_{\mu\nu}\equiv g_{\mu\nu}-\frac{p_{\mu}p_{\nu}}{M_{X_2}^2},
$$
and $\Gamma$ is the $X_2$ width.

It has to be noticed, however, that not all coefficients $C_n$ contribute to the amplitude of the process under study,
when the two-photon tensor $M^{\mu\nu\rho\sigma}$ from Eq.~(\ref{Cndef}) is contracted with the $X_2$ polarisation tensor 
$\varepsilon_{\rho\sigma}^{(\xi)}(p)$ with any polarisation $\xi$. 
Indeed, for convenience, let us define two orthogonal combinations of the basis tensors $e_2^{\mu\nu\rho\sigma}$ and $e_3^{\mu\nu\rho\sigma}$,
namely, $e_2^{\mu\nu\rho\sigma}\sqrt{2}-e_3^{\mu\nu\rho\sigma}$ and $e_2^{\mu\nu\rho\sigma}+\sqrt{2} e_3^{\mu\nu\rho\sigma}$. Then,
using the properties of the $X_2$ polarisation tensor $\varepsilon_{\rho\sigma}^{(\xi)}(p)$ summarised in Eq.~(\ref{tensprop}) 
and the momentum conservation law $p=k_1+k_2$, it is staightforward to see that
\be
e_1^{\mu\nu\rho\sigma} \varepsilon_{\rho\sigma}^{(\xi)}(p)= 
(e_2^{\mu\nu\rho\sigma}+\sqrt{2} e_3^{\mu\nu\rho\sigma}) \varepsilon_{\rho\sigma}^{(\xi)}(p)=0,
\ee
so that only the coefficient $C_4$ and the combination $C_2\sqrt{2}-C_3$ contribute to the amplitude of the two-photon annihilation process which proceeds through
the formation of the tensor state $X_2$.

\subsection{$\gamma\gamma\to D\bar{D}$ annihilation}

The amplitude of the two-photon fusion reaction $\gamma(k_1)\gamma(k_2)\to X_2\to D(p_1)\bar{D}(p_2)$ reads
\be
M(\gamma\gamma\to X_2\to D\bar{D})=g_{X_2D\bar{D}}M^{\mu\nu\rho\sigma}\varepsilon_\mu(k_1)\varepsilon_\nu(k_2){\cal P}_{\rho\sigma,\alpha\beta} 
p_1^\alpha p_2^\beta,
\label{amplDD}
\ee
where $p_1$ and $p_2$ are the momenta of the two $D$ mesons in the final state, respectively, and the $X_2D\bar{D}$ vertex is taken in the form
\be
v_{X_2D\bar{D}}^{\alpha\beta}=g_{X_2D\bar{D}}p_1^\alpha p_2^\beta.
\label{X2DD}
\ee

The kinematics of the process in the centre-of-mass frame is such that ($\venu$ and $\venu'$ are unit vectors)
\be
k_1^\mu=\frac{\sqrt{s}}{2}(1,\venu),\quad k_2^\mu=\frac{\sqrt{s}}{2}(1,-\venu),\quad p_1^\mu=\frac{\sqrt{s}}{2}(1,\eta\venu'),\quad
p_2^\mu=\frac{\sqrt{s}}{2}(1,-\eta\venu'),
\ee
where $\eta=\sqrt{1-4m^2/s}$. Then, the angular distribution is given in terms of $\cos\theta=(\venu\venu')$.

Taking Eqs.~(\ref{Cndef}) and (\ref{amplDD}) together and using the explicit form of the basis tensors (\ref{e1})-(\ref{e4}), it is straightforward 
to find for the differential cross section ($s=p^2=M_{X_2}^2$)
\be
\frac{d\sigma(\gamma\gamma\to X_2\to 
D\bar{D})}{d\cos\theta}=\mbox{const}_{D\bar{D}}\left[\left|A_0\right|^2f_0^{(0)}(\cos\theta)
+2\left|A_{\pm 2}\right|^2f_2^{(0)}(\cos\theta)\right],
\label{xsection}
\ee
where we introduced the helicity-0 and helicity-2 amplitudes,
\be
A_0=C_2\sqrt{2}-C_3,\quad A_{\pm 2}=\sqrt{\frac32}C_4,
\label{A0A2}
\ee
with the same constant, and the quantities $f_0^{(0)}$ and $f_2^{(0)}$ are given by the Wigner functions with $J=2$ and with the helicity equal to 0 and 2, 
respectively, 
\be
f_0^{(0)}(x)={\frac52}[d^2_{0,0}(x)]^2= {\frac58}(3x^2-1)^2 ,\quad f_2^{(0)}(x)={\frac52}[d^2_{2,0}(x)]^2={\frac{15}{16}}(1-x^2)^2.
\ee
Both functions are normalised to unity as
\be
\int_{-1}^1f_0^{(0)}(x)dx=\int_{-1}^1f_0^{(2)}(x)dx=1.
\ee

\subsection{$\gamma\gamma\to \omega J/\psi$ annihilation}

Consider the annihilation reaction $\gamma\gamma\to \omega J/\psi$ accompanied by the subsequent three-pion decay 
$\omega\to \pi^+\pi^-\pi^0$ and by the dilepton decay $J/\psi\to l^+l^-$. The kinematics of the reaction simplifies considerably if one 
notices that, at $\sqrt{s}=M_{X_2}$, the $\omega$ and the $J/\psi$ are quite slow because the annihilation proceeds only about 35~MeV above the 
threshold, which corresponds to a
centre-of-mass momentum $q\approx 200$~MeV. Therefore, neglecting corrections suppressed by the small factors $q/m_\omega$ and $q/m_{J/\psi}$, we consider 
both $\omega$ and $J/\psi$ at rest. 

The angular distribution in $\theta_l$, defined as the angle between the collision axis of the initial photons and the momentum of the 
positively charged lepton $l^+$, follows from the contraction
\be
\frac{d\sigma(\gamma\gamma\to X_2\to 
\omega J/\psi)}{d\cos\theta_l}\sim g_{X_2\omega J/\psi}^2M_{\mu\nu\rho\sigma}^*M^{\mu\nu\rho'\sigma'}{\cal P}^{\rho\sigma,\alpha\beta}{\cal 
P}_{\rho'\sigma',\alpha'\beta'}
N_\alpha^{\alpha'}L_\beta^{\beta'},
\ee
where the tensor $N^{\mu\nu}$ is given by the spectral density of the $\omega$ meson at rest, that is by
\be
N^{\mu\nu}=g^{\mu\nu}-g^{\mu 0}g^{\nu 0},
\ee
and the tensor $L^{\mu\nu}$ is
\be
L^{\mu\nu}=\frac14\mbox{Tr}(\not{p_{l^+}}\gamma^\mu\not{p}_{l^-}\gamma^\nu)=p_{l^+}^\mu p_{l^-}^\nu+p_{l^+}^\nu p_{l^-}^\mu-g^{\mu\nu}(p_{l^+}\cdot 
p_{l^-}),
\ee
with
\be
p_{l^+}^\mu\approx(|\vep_l|,\vep_l),\quad p_{l^-}^\mu\approx(|\vep_l|,-\vep_l), 
\ee
where we neglected the lepton mass as $m_l\ll |\vep_l|=m_{J/\psi}/2$.

It is straightforward then to find that ($s=M_{X_2}^2, \cos\theta_l = (\hat\vep_l\venu)$)
\be
\frac{d\sigma(\gamma\gamma\to X_2\to 
\omega J/\psi)}{d\cos\theta_l}=\mbox{const}_{\omega J/\psi}\left[\left|A_0\right|^2 
f_0^{(1)}(\cos\theta_l)+2\left|A_{\pm 2}\right|^2f_2^{(1)}(\cos\theta_l)\right],
\label{F1}
\ee
where the distribution functions
\be
f_0^{(1)}(x)=\frac18(5-3x^2),\quad f_2^{(1)}(x)=\frac38(1+x^2)
\ee
are normalised to unity,
\be
\int_{-1}^1f_0^{(1)}(x)dx=\int_{-1}^1f_2^{(1)}(x)dx=1,
\ee
and relations (\ref{A0A2}) hold for the helicity amplitudes $A_0$ and $A_{\pm 2}$.
The distribution (\ref{F1}) agrees with the formulae derived from the general principles of the rotational symmetry --- see Eqs.~(12) and (13) of 
Ref.~\cite{Rosner:2004ac}. 

Similarly, the angular distribution in $\theta_n$, defined as the angle between the collision axis of the initial photons and
the normal vector $\ven$ to the plane formed by the three pions originated from the $\omega$ decay ($\cos\theta_n= (\ven\venu$)), follows from 
the contraction
\be
\frac{d\sigma(\gamma\gamma\to X_2\to 
\omega J/\psi)}{d\cos\theta_n}\sim g_{X_2\omega J/\psi}^2M_{\mu\nu\rho\sigma}^*M^{\mu\nu\rho'\sigma'}{\cal P}^{\rho\sigma,\alpha\beta}{\cal 
P}_{\rho'\sigma',\alpha'\beta'}\tilde{N}_\alpha^{\alpha'}\tilde{L}_\beta^{\beta'},
\ee
where the tensor $\tilde{N}^{\mu\nu}$ is now defined as
\be
\tilde{N}^{\mu\nu}=q^\mu q^\nu,\quad q^\mu=(0,\ven),
\ee
and the tensor $\tilde{L}^{\mu\nu}$ is given by the spectral density of the $J/\psi$ at rest, that is by
\be
\tilde{L}^{\mu\nu}=g^{\mu\nu}-g^{\mu 0}g^{\nu 0}.
\ee

Then, at $s=M_{X_2}^2$,
\be
\frac{d\sigma(\gamma\gamma\to X_2\to 
\omega J/\psi)}{d\cos\theta_n}=\mbox{const}_{\omega J/\psi}\left[\left|A_0\right|^2 
f_0^{(2)}(\cos\theta_n)+2\left|A_{\pm 2}\right|^2f_2^{(2)}(\cos\theta_n)\right],
\label{F2}
\ee
where, as before, the same coefficients from Eq.~(\ref{A0A2}) appear and the distribution functions
\be
f_0^{(2)}(x)=\frac14(1+3x^2),\quad f_2^{(2)}(x)=\frac34(1-x^2)
\ee
are normalised to unity,
\be
\int_{-1}^1f_0^{(2)}(x)dx=\int_{-1}^1f_2^{(2)}(x)dx=1.
\ee

Distribution (\ref{F2}) agrees with Eqs.~(12) and (14) of Ref.~\cite{Rosner:2004ac}.

\subsection{Evaluation of the helicity amplitudes}

From Eqs.~(\ref{xsection}), (\ref{F1}), and (\ref{F2}) one can see that, in agreement with the natural expectations, the angular distribution takes a 
universal form,
\be
\frac{d\sigma(\gamma\gamma\to X_2\to \mbox{final 
state})}{d\cos\theta}=\mbox{const}_{f}\left[|A_0|^2f_0^{(f)}(\cos\theta)+2|A_{\pm 2}|^2f_2^{(f)}(\cos\theta)\right],
\label{F}
\ee
where the functions $f_0^{(f)}$ and $f_2^{(f)}$ depend on the particular final state, as derived
above, and are given by the normalised helicity-0 and helicity-2 distribution, 
respectively~\cite{Rosner:2004ac,Zhou:2015uva}. Similarly, $\mbox{const}_f$ in Eq.~(\ref{F}) stands for an overall constant depending on the final state in the two photon fusion process. 
The dynamics is encoded in the helicity amplitudes
$A_0$ and $A_{\pm 2}$ defined in Eq.~(\ref{A0A2}).
Therefore, one can conclude that the tensor structures (\ref{e1})-(\ref{e3}) correspond to the helicity equal to 0
while the tensor structure (\ref{e4}) corresponds to the helicity equal to 2. 
The form of the angular distribution given by Eq.~(\ref{F}) depends on the relative strength of the coefficients, 
and we may define the ratio
\be
R\equiv \frac{2|A_{\pm 2}|^2}{|A_0|^2}
\label{ratioR}
\ee
 to rewrite the differential cross section (\ref{F}) in the form
\be
\frac{d\sigma(\gamma\gamma\to X_2\to \mbox{final 
state})}{d\cos\theta}=\sigma_0(\gamma\gamma\to X_2\to \mbox{final 
state})\left[f_0^{(f)}(\cos\theta)+R\,f_2^{(f)}(\cos\theta)\right],
\label{FR}
\ee
where $\sigma_0$ corresponds to the total cross section evaluated solely for the helicity-0 amplitude.

An explicit evaluation of the loop integrals yields that 
their contributions to the coefficients $\{C_n\}$ are divergent and thus call for an
additional counter terms to render the amplitude well defined. 
However, due to the property $g_{\rho\sigma}X_2^{\rho\sigma}=0$ (see Eq.~(\ref{tensprop})), the structure (\ref{S1}) 
and correspondingly the coefficient $C_1$ does 
not contribute to observables and, as was demonstrated above, the structures (\ref{S2}) and (\ref{S3}) contribute to the 
$\gamma\gamma\to X_2$ transition amplitude in the combination $C_2\sqrt{2}-C_3$ which is 
\emph{finite} --- see the Appendix~\ref{app:hel0} for the explicit form of the amplitude $A_0$.
This observation suggests that local counterterms for this contribution appear only at a higher order. 
In particular, the leading-order short-range contributions to the two-photon decay proceeding 
via excitation of two vector mesons in the intermediate state (for example, $\omega J/\psi$) or via the quarkonium component of the $X_2$ wave function 
contribute to the helicity-2 amplitude only.
Therefore, once the $X_2D^*\bar D^*$ coupling is fixed via its connection to the $X(3872)$ coupling (see Eqs.~(\ref{couplings}) below), 
the helicity-0 contribution comes as a 
prediction of the model. On the other hand, to quantify the helicity-2 amplitude, a counter term needs to be fixed 
by some data. 
This shows that the relative importance of the two helicity components
is, in general, not fixed by the structure. On the other hand, as 
will be shown below, for a vanishing binding energy 
the contribution of the helicity-0 amplitude
vanishes, and a helicity-2 dominance in the observables appears
to be natural, in analogy to regular charmonia. 

In order to proceed we now calculate the helicity-0 component within the molecular model outlined 
and then use data to fix the ratio $R$ introduced in Eq.~(\ref{ratioR}).

\section{Ratio of the helicity amplitudes from data}

\subsection{The coupling $X_2\to D^*\bar{D}^*$}
\label{sec:X2D*D*}

Under the assumption that the $X_2$ is a $D^*\bar{D}^*$ molecule, its dominating decay mechanism proceeds through the $S$-wave $X_2\to D^*\bar{D}^*$
vertex followed by $D^{(*)}$-meson loops. In order to estimate the coupling constant $g_{X_2D^*\bar{D}^*}$, we 
employ the assumption that the $X_2$ is a spin partner of the $X(3872)$ and, therefore, they both
appear as members of the same superfield $\chi^i$ --- see, for example, Ref.~\cite{Mehen:2015efa},
\bea
\chi^i=\sigma^j\chi^{ij}=\sigma^j\left(\chi^{ij}_2+\frac{1}{\sqrt{2}}\epsilon^{ijk}\chi^k_1+\frac{\delta^{ij}}{\sqrt{3}}\chi_0\right),
\label{chi}
\eea
where the nonrelativistic fields $\chi^{ij}_2$, $\chi^i_1$, and $\chi_0$ describe the $X_2$, the $X_1\equiv X(3872)$, and the hypothetical 
scalar molecular state $X_0$, 
respectively. 
The field (\ref{chi}) interacts with the vector and pseudoscalar heavy-light mesons as
\be
\Lag_\chi=i\frac{g_1}{2}{\rm Tr}[\chi^{\dagger \, i} H_a \sigma^i\Hb_a],
\label{lagchi}
\ee
where
\be
H_a =\veV_a \cdot\vesig+P_a,\quad \bar{H}_a =-\bar{\veV}_a \cdot\vesig+\bar{P}_a,
\label{Hfield}
\ee
with $\veV_a$ ($\bar{\veV}_a$) and $P_a$ ($\bar{P}_a$) denoting the vector and pseudoscalar meson (antimeson) states, respectively. 

Substituting Eqs.~(\ref{chi}) and (\ref{Hfield}) into Eq.~(\ref{lagchi}) 
and taking traces, one gets relations between the nonrelativistic couplings consistent with HQSS,
\be
g^{\rm nr}_{X_0V\bar{V}}=\frac{1}{\sqrt{3}}g_1,\quad 
g^{\rm nr}_{X_1P\bar{V}}=g^{\rm nr}_{X_1V\bar{P}}=g^{\rm nr}_{X_2V\bar{V}}=2g_1,
\label{couplings}
\ee
where, in our calculation, $P$ and $V$ correspond to the $D$ and $D^*$ meson, 
respectively.\footnote{Note that due to the $D$-wave character of the vertex defined in Eq.~(\ref{X2DD})
the coupling $g_{X_2D\bar{D}}$ is not related to the couplings that appear in Eq.~(\ref{couplings}).}

To proceed, we use the $X(3872)$ state as the anchor since, in the molecular scenario, its coupling constant to the $D\bar{D}^*$ pair 
$g^{\rm nr}_{X_1D\bar{D}^*}$ can be evaluated through its measured binding energy $E_B$ as \cite{Guo:2013nza} 
(see also Refs. \cite{Landau,Weinberg:1965zz,Baru:2003qq} for more general discussions)
\be
\left(g^{\rm nr}_{X_1D\bar{D}^*}\right)^2\approx \frac{16\pi}{\mu}\sqrt{\frac{2E_B}{\mu}}+\ldots,\quad \mu=\frac{mm_*}{m+m_*},
\label{gmu}
\ee
where $m$ and $m_*$ stand for the $D$ and $D^*$-meson masses, respectively,\footnote{Strictly speaking, in the HQSS limit, the $D$ and $D^*$ mesons
are degenerate in mass, so that one cannot distinguish between $m$ and $m_*$. For definiteness, up to the corrections to 
the heavy-quark limit, we define the reduced mass $\mu$ as given in 
Eq.~(\ref{gmu}) and use it in the other coupling constants as well --- see Eq.~(\ref{gmu2}) below.} and 
the ellipsis denotes the terms suppressed as $\sqrt{\mu E_B}/\beta$, with $\beta$ being the inverse range of the force.
Thus, the nonrelativistic coupling constants for the scalar and tensor $D^*\bar{D}^*$ spin partners of the $X(3872)$ read 
\be
\left(g^{\rm nr}_{X_0D^*\bar{D}^*}\right)^2=\frac{4\pi}{3\mu}\sqrt{\frac{2E_B}{\mu}},\quad
\left(g^{\rm nr}_{X_2D^*\bar{D}^*}\right)^2=\left(g^{\rm nr}_{X_1D\bar{D}^*}\right)^2=\frac{16\pi}{\mu}\sqrt{\frac{2E_B}{\mu}}.
\label{gmu2}
\ee

Then the interaction Lagrangian with relativistic vertices can be written as 
\be
\Lag=ig_{X_0D^*\bar{D}^*}X_0^{\dag}D^{*\mu}\bar{D}^*_\mu
-\frac{1}{\sqrt{2}}g_{X_1D^*\bar{D}^*}X_{1\mu}^{\dag}\left(D^{*\mu}\bar{D}+D\bar{D}^{*\mu}\right)
-ig_{X_2D^*\bar{D}^*}X_{2\mu\nu}^{\dag}D^{*\mu}\bar{D}^{*\nu}
+\mbox{h.c.},
\label{lagXnr}
\ee
where 
\bea\label{couplingsrel}\nonumber
\frac{g^2_{X_0D^*\bar{D}^*}}{4\pi}&=&\frac{M_{X_0}}{3m}(m+m_*)^{3/2}\sqrt{\frac{2m_*E_B}{m}},\nonumber\\[-2mm]\nonumber
\\[-2mm]
\frac{g^2_{X_2D^*\bar{D}^*}}{4\pi}
&=&\frac{4M_{X_2}}{m}(m+m_*)^{3/2}\sqrt{\frac{2m_*E_B}{m}},\\[-2mm] \nonumber\\[-2mm]
\frac{g^2_{X_1D\bar{D}^*}}{4\pi}&=&\frac{4M_{X_1}}{m_*}(m+m_*)^{3/2}\sqrt{\frac{2m_*E_B}{m}},\nonumber
\eea
and $M_{X_J}$ ($J=0,1,2$) stands for the mass of the resonance with the total spin $J$. To arrive at Eq.~\eqref{couplingsrel}, 
we used the relation between the relativistic and nonrelativistic coupling constants, namely 
\be
g_{m_1\to m_2m_3}=\sqrt{m_1m_2m_3}\;g^{\rm nr}.
\ee

\subsection{Two-photon decay width and the ratio of the helicity amplitudes}
\label{Sec:GX2gg}

Consider the two-photon decay of the $X_2$ molecule which, due to the $T$-invariance, is described by the amplitude (\ref{amplitude1}). 
Then the differential width can be written 
in a form similar to Eq.~(\ref{FR}). Since the angular distributions $f_0^{(f)}$ and $f_2^{(f)}$ are normalised to unity, 
the angular integration yields
\be
\Gamma(X_2\to\gamma\gamma)=\Gamma_0(X_2\to\gamma\gamma)\left[1+R\right],
\label{GR}
\ee
where, as before, $\Gamma_0(X_2\to\gamma\gamma)$ denotes the width evaluated solely for the helicity-0 amplitude.
We stress again that, within the model at hand, this piece is fixed in terms of the coupling constant of the
$X(3872)$ to $D\bar{D}^*$.

Using the explicit form of the amplitude (\ref{Cndef}) and the value of the coupling constant $g_{X_2D^*\bar{D}^*}$ from Eq.~(\ref{couplingsrel}), 
one can find that (the $X(3872)$ binding energy $E_B$ is measured in MeV)
\be
\Gamma_0(X_2\to\gamma\gamma)\approx 0.033\sqrt{E_B}~\mbox{keV},
\label{G0}
\ee
which comes out as a prediction of the model used.\footnote{The calculations were performed with the help of 
Wolfram Mathematica supplied with the FeynCalc \cite{Mertig:1990an,Shtabovenko:2016sxi} and Package-X \cite{Patel:2015tea} packages. 
In addition, for a cross check, the LoopTools package was employed \cite{Hahn:1998yk}.}

In order to proceed, one needs an estimate of the $X(3872)$ binding energy. 
The most up-to-date value of the $X$ mass reads \cite{Olive:2016xmw}
\be
M_{X_1}=(3871.69\pm 0.17)~\mbox{MeV},
\ee
which gives for the $E_B$
\begin{eqnarray}
E_B=m_0+m_{*0}-M_{X_1}=0.01\pm 0.20~\mbox{MeV}.
\label{EXexp} 
\end{eqnarray}

Since the central value of the $E_B$ is basically consistent with zero, for an estimate, we take the upper bound $E_B=0.21$~MeV. Then, 
Eq.~(\ref{G0}) yields
\be
\Gamma_0(X_2\to\gamma\gamma)\lesssim 0.015~\mbox{keV}.
\label{G0num}
\ee
We may now employ this result in Eq.~(\ref{GR}) to extract the ratio $R$ from the data on $\Gamma(X(3915)\to\gamma\gamma)$.
Unfortunately, the existing data are rather uncertain and, in addition, the analysis may implicitly include a particular assumption about the quantum 
numbers of the state 
and/or about a particular helicity dominance for it, like it was done in Ref.~\cite{Lees:2012xs}. We, therefore, 
consider several results of different experimental measurements simultaneously to reliably estimate
the two-photon decay width $\Gamma(X(3915)\to\gamma\gamma)$.

We start from the Belle result \cite{Uehara:2005qd}~\footnote{While Belle calls this state $Z(3930)$ we here
employ the PDG naming scheme and call it $\chi_{c2}(3930)$}
\be
\Gamma(\chi_{c2}\to\gamma\gamma){\cal B}(\chi_{c2}\to D\bar{D})=(0.18\pm 0.05\pm 0.03)~\mbox{keV}
\ee
and, following Ref.~\cite{Zhou:2015uva}, identify the tensor state $\chi_{c2}(3930)$ with $X_2(3915)$. Then we use the fact that
its total width is largely saturated by the $D\bar{D}$ mode, so, for an estimate, it is natural to 
set ${\cal B}(\chi_{c2}(2P)\to D\bar{D})\approx 1$ \cite{Olive:2016xmw}. This leads to the estimate
\be
\Gamma(X(3915)\to\gamma\gamma)\simeq 0.18~\mbox{keV}.
\label{widthexp}
\ee

An alternative way to extract the $X(3915)$ two-photon width is to use the measurements of the $\Gamma(X(3915)\to\gamma\gamma){\cal B}(X(3915)\to\omega J/\psi)$. In particular,
the Belle result from Ref.~\cite{Uehara:2009tx} reads
\be
\Gamma(X(3915)\to\gamma\gamma){\cal B}(X(3915)\to\omega J/\psi)= 
\left\{ 
\begin{array}{ll}
(61 \pm 17 \pm 8)~ {\rm eV}, & J^P=0^+ \\
(18 \pm 5 \pm 2)~ {\rm eV}, & J^P=2^+, 
\end{array} \right.
\label{expwidth1}
\ee
while BaBar gives \cite{Lees:2012xs}
\be
\Gamma(X(3915)\to\gamma\gamma){\cal B}(X(3915)\to\omega J/\psi)= 
\left\{ 
\begin{array}{ll}
(52 \pm 10 \pm 3)~ {\rm eV}, & J^P=0^+ \\
(10.5 \pm 1.9 \pm 0.6)~ {\rm eV}, & J^P=2^+.
\end{array} \right.
\label{expwidth2}
\ee
Thus, for the estimate, we take an averaged value
\be
\Gamma(X_2\to\gamma\gamma){\cal B}(X(3915)\to\omega J/\psi)\approx 15~{\rm eV},
\label{expwidth}
\ee
which arises if the above measurements for the quantum numbers $J^P=2^+$ are used. 

Now, to extract the width $\Gamma(X_2\to\gamma\gamma)$ from Eq.~(\ref{expwidth}), we use the experimental value for the product of the BF's 
\be
{\cal B}(B\to X(3915)K){\cal B}(X(3915)\to\omega J/\psi)=(0.51\pm 0.11)\cdot 10^{-4},
\label{Bav}
\ee
which is quoted by the PDG \cite{Olive:2016xmw} as the average between the BaBar and Belle measurements --- see 
Refs.~\cite{delAmoSanchez:2010jr} and \cite{Abe:2004zs}, respectively. The value (\ref{Bav}) is also consistent with another BaBar measurement reported in 
Ref.~\cite{Aubert:2007vj}.

If, in addition, we assume that the $X(3915)$ production branching falls into the ball park of the BF's for the 
production of other known charmonia in $B$-meson weak decays \cite{Olive:2016xmw}, that is, that it constitutes a few units times $10^{-4}$, then 
it is straightforward to arrive at the estimate
\be
{\cal B}(X(3915)\to\omega J/\psi)\simeq (10\div 20)\%,
\ee
which allows one to extract the width $\Gamma(X(3915)\to\gamma\gamma)$ from (\ref{expwidth}) to be
\be
\Gamma(X(3915)\to\gamma\gamma)\simeq 0.1~\mbox{keV}
\label{widthexp2}
\ee
which is fairly consistent with the independent estimate of Eq.~(\ref{widthexp}). 

As an additional consistency check we notice that, starting from the two-photon decay width of the $\chi_{c2}(1P)$ charmonium \cite{Ablikim:2012xi},
\be
\Gamma(\chi_{c2}(1P)\to\gamma\gamma)=(0.66\pm 0.07\pm 0.06)~\mbox{keV},
\ee
and taking into account that it is natural to expect a few times smaller value of the width for the excited tensor state, one qualitatively re-arrives at the estimate
(\ref{widthexp}). 

\begin{figure}[t]
\begin{center}
\epsfig{file=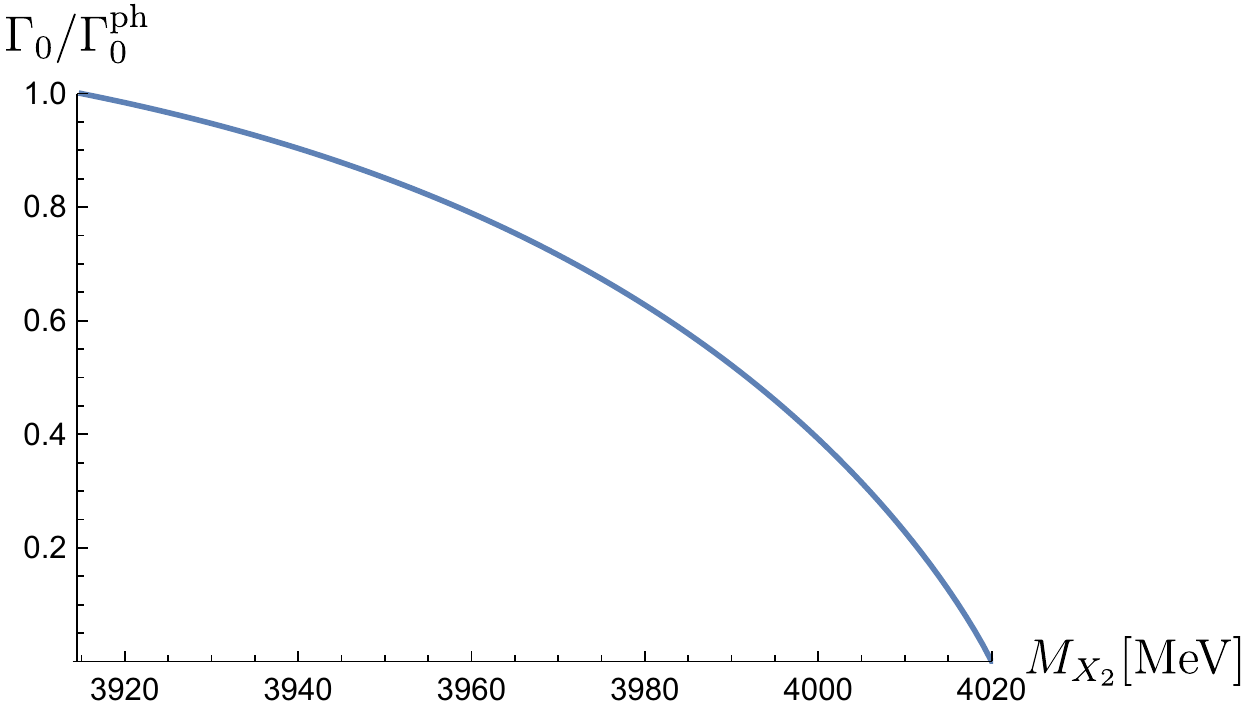,width=0.6\textwidth}\hspace*{5mm}
\end{center}
\caption{The helicity-0 contribution to the width, $\Gamma_0(X_2\to\gamma\gamma)$, in units of its physical value
(that is, its value at the mass 3915 MeV) as a function of the mass of the $X_2$ state. In this ratio, the binding energy $E_B$ is cancelled.}\label{GaXM}
\end{figure}

Now, with the theoretical estimate of the helicity-0 contribution given in Eq.~(\ref{G0num}) and with the 
phenomenological estimate for the total two-photon width of Eq.~(\ref{widthexp}) at hand, we are in a position to estimate the value 
of the ratio $R$ employing Eq.~(\ref{GR}). This yields
\be
R\gtrsim 11\gg 1,
\label{R}
\ee
where, as it was explained above, to arrive at this estimate we used the upper bound on the $X(3872)$ binding energy $E_B$. In the limit $E_B\to 0$, the ratio $R$ 
grows as $1/\sqrt{E_B}$. 

\section{Discussion, Disclaimers and possible future improvements}
\label{sec:DDfi}

The result of Eq.~(\ref{R}) suggests a strong helicity-2 amplitude dominance in the two-photon decay of 
a 2$^{++}$ $D^*\bar{D}^*$ molecular state in the mass range of 3915 MeV under the assumption that this state is the spin partner of the $X(3872)$,
the latter assumed to be a $D\bar{D}^*$ bound state. 
Moreover, as one can see from Fig.~\ref{GaXM}, 
had the mass of the $X_2$ state been located closer to the $D^*\bar D^*$ threshold, strong cancellations between 
the contributions to the helicity-0 amplitude from the diagrams depicted in Fig.~\ref{fig:diagramsmagn} with the intermediate $D^*DD^*$ and 
$D^*D^*D^*$ particles
would have taken place thus damping the helicity-0 contribution to the two-photon width. As a consequence, 
the state would have been almost entirely dominated by the 
helicity-2 amplitude, unless it had completely decoupled from $\gamma\gamma$.
Thus the behaviour were analogous to that of Ref.~\cite{Li:1990sx} for the genuine charmonium assignment and Refs.~\cite{Alekseev,Krammer:1977an} for the positronium.
On the other hand, as shown in Ref.~\cite{Zhou:2015uva}, the angular distributions measured by BaBar are consistent with the $X(3915)$ being either a tensor state with 
the large helicity-0 contribution or a scalar state. Based on the analysis presented in this paper, one is, therefore,
tempted to conclude that if the $X(3915)$ is indeed the helicity-0 contribution of the nearby tensor state, which requires an exotic structure
of the latter, it is not the spin partner of the $X(3872)$.

However, there were various assumptions used to arrive at the estimate of Eq.~(\ref{R}). First of all, there are quite some uncertainties in the
estimate for $\Gamma(X(3915)\to\gamma\gamma)$ given in Eq.~(\ref{widthexp}). Clearly, this can be improved by more refined measurements
of this quantity. Next, in order to employ Eq.~(\ref{gmu}), we needed to assume the $X(3872)$ to be a $D\bar{D}^*$ bound state. 
However, current data on the $X(3872)$ are also compatible with a virtual state~\cite{Hanhart:2007yq} and then Eq.~(\ref{gmu}),
derived from the normalisation of the bound state wave function, does not hold anymore. In the case of a virtual state, the extraction of the coupling constant 
could be done directly by fitting the experimental line shapes along the lines of Refs.~\cite{Hanhart:2007yq,Kang:2016jxw}, which, however, requires 
high-resolution and high-statistics data not available at present. In this context, it is also important to remember that the data currently available
for $X(3872)\to D\bar{D}^*$ channel were subject to a kinematic fit that might have distorted the line shape~\cite{Stapleton:2009ey}.

In order to relate the coupling of the $X(3872)$ to that of its spin partner $X_2$ we needed to use heavy-quark spin symmetry.
However, in the charm sector, there might occur a significant spin symmetry violation, since $\Lambda_{\rm QCD}/m_c\sim 0.2$ is not
a small parameter. Furthermore, especially for the $X_2$, there might be significant spin symmetry violating contributions induced by the 
sizeable mass difference of the $D\bar{D}$ and $D^*\bar{D}^*$ intermediate states~\cite{Baru:2016iwj}. In order to improve on this
piece of the calculation, a better understanding of the impact of spin symmetry violations on spin multiplets is required.
In addition, from the theoretical side, the role of a mixing with $\bar cc$ states needs to be better understood~\cite{Cincioglu:2016fkm} --- see
also Ref.~\cite{Hammer:2016prh} where this issue is approached from a different angle.
Clearly, these are major tasks that need both further calculations as well as further data, especially for additional
quantum numbers~\cite{Cleven:2015era}, not only in the charm sector but also in the bottom sector~\cite{Bondar:2016hva}.
In this sense, the study presented in this work can be regarded as an additional step towards a concise and complete understanding
of the heavy meson spectrum above the open-flavour threshold.

\section{Conclusions}

Even almost 15 years after the discovery of the $X(3872)$ the heavy-meson spectrum above the open-flavour
thresholds still raises a lot of questions. Many proposals are put forward to explain the large number of
unusual states discovered in the mean time and so far no coherent picture is on the horizon. 
In order to improve the situation, it appears necessary to better understand the implications of the
heavy-quark spin symmetry and its violations on the spectrum of heavy hadrons which are very different
in the different scenarios~\cite{Cleven:2015era}. 

In this paper, we treat the $X(3915)$ as a $D^*\bar{D}^*$ molecule to investigate if within this scenario it could indeed 
be the tensor state known as $\chi_{c2}(3930)$ --- an identification
that calls for a prominent helicity-0 component in the wave function of the state~\cite{Zhou:2015uva}. 
To this end, we study the two-photon
annihilation processes proceeding through the $X(3915)$ and evaluate the two-photon decay width of this state. 
We find that, for such a molecule, the helicity-0 component vanishes as
the binding energy tends to zero while the helicity-2 piece (which contains an unknown counter term) 
is expected to be finite, unless the $X(3915)$ completely decouples from the $\gamma\gamma$ channel in this limit.
Thus, it appears natural that shallow bound states share
the feature of a helicity-2 dominance with regular charmonia.
However, the state at hand is located about 100 MeV below the $D^*\bar{D}^*$ threshold and, therefore,
no \emph{a priori} conclusion about the relative importance of the helicity amplitudes is possible.

In order to proceed, we investigate whether or not the $X(3915)/\chi_{c2}(3930)$ could be the spin-2 partner of the $X(3872)$
which is assumed to be a $D\bar{D}^*$ bound state.
To this end, we evaluate the contribution of the helicity-0 
amplitude to the two-photon decay width of
the $X(3915)$, which is finite and comes out as a prediction of the
model once spin symmetry is used to connect the effective coupling constant
of the $X(3915)$ to that of the $X(3872)$.
We argue that the experimental data presently available for the $X(3872)$ mass and for the two-photon decay width of the tensor charmonium 
favour a scenario in which the contribution of the helicity-2 amplitude dominates over the helicity-0 one,
similarly to the case of the genuine $2^{++}$ charmonium. 

In summary, if the $X(3915)$ is the $D^{(*)}\bar D^{(*)}$ hadronic molecule that should exist as a spin partner of the $X(3872)$, current
data favour a scalar assignment for this state. On the other hand, if
the $X(3915)$ were indeed dominated by the helicity-0 contribution of the nearby tensor state, its nature would require some exotic 
interpretation related neither with regular quarkonia nor with the $D^*\bar{D}^*$ spin partner of the $X(3872)$.
However,
it needs to be acknowledged that the analysis performed
is subject to several uncertainties which are difficult to quantify, given the present status of the experimental data
as well as our theoretical understanding of the charmonium spectrum above the open flavour thresholds. 
We also outline how those uncertainties could be reduced in future studies.

\acknowledgments

The authors are grateful to F.-K. Guo and to R. V. Mizuk for enlightening discussions. 
This work is supported in part by the DFG and the NSFC through funds provided to the Sino-German CRC 110 ``Symmetries 
and the Emergence of Structure
in QCD'' (NSFC Grant No. 11261130311). Work of A. N. was performed within the Institute of Nuclear Physics and Engineering supported by 
MEPhI Academic Excellence Project (contract No 02.a03.21.0005, 27.08.2013). He also acknowledges support from the
Russian Foundation for Basic Research (Grant No. 17-02-00485). 
Work of V. B. is supported by the DFG (Grant No. GZ: BA 5443/1-1). 

\appendix

\section{Helicity-0 amplitude}\label{app:hel0}

In this Appendix we collect explicit expressions for the contributions $C_2[D^*D^{(*)}D^*]$ and $C_3[D^*D^{(*)}D^*]$ 
defined in Eq.~(\ref{Cndef}) (here the particle which appears in the middle propagates 
between the photon emission vertices) to the helicity-0 amplitude $A_0[D^*D^{(*)}D^*]$ from Eq.~(\ref{A0A2}) 
which stem from the triangle diagrams depicted in Fig.~\ref{fig:diagramsmagn}. The individual contributions read (here $M\equiv M_{X_2}$)
\bea
&&C_2[D^*DD^*]=\frac{1}{16\pi^2\sqrt2}{\left(a_1 L \sqrt{1-\frac{4m_*^2}{M^2}}+2M^2a_2\text{C}_0\left(0,0,M^2,m_*^2,m^2,m_*^2\right)+a_3\right)}, 
\nonumber\\[2mm]
&&C_3[D^*DD^*]=0, 
\\[2mm]
&&C_2[D^*D^*D^*]=\frac{1}{16\pi^2\sqrt2}{\left( { b_0} \lambda+b_1 L 
\sqrt{1-\frac{4m_*^2}{M^2}}+2M^2b_2\text{C}_0\left(0,0,M^2,m_*^2,m_*^2,m_*^2\right)+b_3\right)},\nonumber
\\[2mm]
&&C_3[D^*D^*D^*]=\frac{1}{16\pi^2}{\left(b_0\lambda+b_4 L 
\sqrt{1-\frac{4m_*^2}{M^2}}+2M^2b_5\text{C}_0\left(0,0,M^2,m_*^2,m_*^2,m_*^2\right)+b_6\right)},\nonumber
\eea
where $\lambda=-1/\epsilon-\log\left(\mu^2/m_*^2\right)-\log(4\pi)+\gamma_E-1$, $\mu$ stands for the renormalisation scale in the 
dimensional regularisation scheme, and $\gamma_E\approx 0.577$ is Euler's constant. 
Furthermore, \\ $L=\ln{\left(\frac{-M^2+\sqrt{M^4-4 M^2 m_*^2}+2m_*^2}{2 m_*^2}\right)}$, $\text{C}_0$ is the scalar Passarino-Veltman function, which is finite
(for example, $\text{C}_0\left(0,0,M^2,m_*^2,m_*^2,m_*^2\right)=L^2/(2M^2)$), and the coefficients $a_n$ and $b_n$ read
\begin{align}
&a_1=-\left(m^2-m_*^2\right)\text{Tr}_{D}^{mm},\nonumber\\
&a_2=-\frac14\left(m^2+\frac{2(m^2-m_*^2)^2}{M^2}\right)\text{Tr}_{D}^{mm},\nonumber\\
&a_3=-\left(m^2-m_*^2-m^2\ln{\left(\frac{m_*^2}{m^2}\right)}+\frac14M^2\right)\text{Tr}_{D}^{mm},\nonumber\\
&b_0=\frac{M^4 \left(\text{Tr}_{D^*}^{ee}-m_*^2 \text{Tr}_{D^*}^{mm}\right)}{12 m_*^4},\nonumber\\
&b_1=\frac{M^2 \text{Tr}_{D^*}^{em}}{m_*}+\text{Tr}_{D^*}^{mm} \left(\frac{M^4}{12 m_*^2}-\frac{M^2}{3}\right)
+\text{Tr}_{D^*}^{ee} \left(-\frac{M^4}{12 m_*^4}+\frac{4 M^2}{3m_*^2}+4\right),\\
&b_2=\frac{M^2 \text{Tr}_{D^*}^{em}}{4 m_*}+\frac{M^2 \text{Tr}_{D^*}^{mm}}{8}+\text{Tr}_{D^*}^{ee} \left(1
+\frac{M^2}{8 m_*^2}+\frac{2 m_*^2}{M^2} \right),\nonumber\\
&b_3=\text{Tr}_{D^*}^{em} \left(\frac{M^4}{8 m_*^3}+\frac{2 M^2}{m_*}\right)+\text{Tr}_{D^*}^{mm} 
\left(\frac{17 M^4}{144 m_*^2}-\frac{2 M^2}{3}\right)+\text{Tr}_{D^*}^{ee} \left(\frac{M^4}{144 
m_*^4}+\frac{19 M^2}{6 m_*^2}+10\right),\nonumber\\
&b_4=-\frac{M^2 \text{Tr}_{D^*}^{em}}{2 m_*}+\text{Tr}_{D^*}^{mm} \left(\frac{M^4}{12 m_*^2}-\frac{7 M^2}{12}\right)+\text{Tr}_{D^*}^{ee} 
\left(-\frac{M^4}{12 m_*^4}+\frac{M^2}{12 m_*^2}-2\right),\nonumber\\
&b_5= -m_* \text{Tr}_{D^*}^{em}-\text{Tr}_{D^*}^{ee},\nonumber\\
&b_6=-\text{Tr}_{D^*}^{em} \left(\frac{2 M^2}{m_*}-\frac{M^4}{12 m_*^3}\right)+\text{Tr}_{D^*}^{mm} \left(\frac{7 M^4}{72 m_*^2}-\frac{7 M^2}{6}\right)+\text{Tr}_{D^*}^{ee} \left(-\frac{M^4}{72 
m_*^4}-\frac{M^2}{2 m_*^2}-4\right).\nonumber
\end{align}
Here $\text{Tr}_D^{mm}$, $\text{Tr}_{D^*}^{mm}$, $\text{Tr}_{D^*}^{em}$, and $\text{Tr}_{D^*}^{ee}$ stand for the isospin traces taken for the 
triangle diagram 
with the $D^{(*)}$ meson propagator between the photon emission vertices. The superscripts $mm$, $em$, and $ee$ denote two magnetic, one magnetic 
plus one electric, and two electric photon emission vertices, respectively. Then, with the help of the explicit form of the matrices 
$\hat{\cal M}_{D^*D^{(*)}}$ 
(see Eq.~(\ref{Mmatr})) and the matrix $\hat{Q}_D$ (see below Eq.~(\ref{psipsigam})) one readily finds
\bea
\text{Tr}_D^{mm}&=&\Tr\hat{\cal M}_{D^*D}^2=\frac{5\beta^2}{9}+\frac{4\beta}{9m_c}+\frac{8}{9m_c^2},\nonumber\\
\text{Tr}_{D^*}^{mm}&=&\Tr\hat{\cal M}_{D^*D^*}^2=\frac{5\beta^2}{9}-\frac{4\beta}{9m_c}+\frac{8}{9m_c^2},\nonumber\\[-2mm]
\\[-2mm]
\text{Tr}_{D^*}^{em}&=&\Tr\left(\hat{\cal M}_{D^*D^*}\hat{Q}_D\right)=-\frac{\beta}{3}-\frac{2}{3 m_c},\nonumber\\
\text{Tr}_{D^*}^{ee}&=&\Tr\hat{Q}_D^2=1.\nonumber
\eea

We emphasise once more that while both functions $C_2[D^*D^*D^*]$ and $C_3[D^*D^*D^*]$ diverge, their combination
$C_2\sqrt{2}-C_3$ which contributes to $A_0$ is finite
\bea
A_0[D^*DD^*]&=&\frac{1}{16\pi^2}{\left(a_1 L \sqrt{1-\frac{4m_*^2}{M^2}}+2M^2a_2\text{C}_0\left(0,0,M^2,m_*^2,m^2,m_*^2\right)+a_3\right)},\nonumber\\[-2mm]
\\[-2mm]
A_0[D^*D^*D^*]&=&\frac{1}{16\pi^2}{\left(c_1 L \sqrt{1-\frac{4m_*^2}{M^2}}+2M^2c_2\text{C}_0\left(0,0,M^2,m_*^2,m_*^2,m_*^2\right)+c_3\right)}\nonumber,
\eea
where the coefficients $c_n$ read $c_1=b_1 -b_4,\ \ c_2=b_2-b_5$ and $c_3=b_3-b_6$.

\section{Helicity-2 amplitude}\label{app:hel2}

For completeness, in this Appendix we collect explicit expressions for the contributions $C_4[D^*D^{(*)}D^*]$ 
defined in Eq.~(\ref{Cndef}) (as in Appendix~\ref{app:hel0}, in this notation, the particle which appears in the middle propagates 
between the photon emission vertices) to the helicity-2 amplitude $A_{\pm2}[D^*D^{(*)}D^*]$ from Eq.~(\ref{A0A2}) 
which stem from the triangle diagrams depicted in Fig.~\ref{fig:diagramsmagn}. The individual contributions are (here $M\equiv M_{X_2}$)
\bea
&&C_4[D^*DD^*]=\frac{1}{32\pi^2\sqrt2}{\left( { d_0} \lambda+d_1 L 
\sqrt{1-\frac{4m_*^2}{M^2}}+2M^2d_2\text{C}_0\left(0,0,M^2,m_*^2,m^2,m_*^2\right)+d_3\right)},\nonumber
\\[2mm]
&&C_4[D^*D^*D^*]=\frac{1}{16\pi^2\sqrt2}{\left(e_0\lambda+e_1 L 
\sqrt{1-\frac{4m_*^2}{M^2}}+2M^2e_2\text{C}_0\left(0,0,M^2,m_*^2,m_*^2,m_*^2\right)+e_3\right)},\nonumber
\eea
and the coefficients $d_n$ and $e_n$ read
\begin{align}
&d_0={\text{Tr}_D^{mm}\, M^2 },\nonumber\\
&d_1=-{\text{Tr}_D^{mm} \left(-2 m^2+M^2+2 m_*^2\right)},\\
&d_2=\frac{\text{Tr}_D^{mm} \left(m^2-m_*^2\right)^2}{M^2},\nonumber\\
&d_3={\text{Tr}_D^{mm} \left(-2 m^2 \ln{\left(\frac{m_*^2}{m^2}\right)}+2 m^2+M^2-2 m_*^2\right)},\nonumber\\
&e_0=\frac{M^4 \text{Tr}_{D^*}^{em}}{12 m_*^3}+\text{Tr}_{D^*}^{mm} \left(\frac{M^4}{24 m_*^2}+{M^2}\right)+\text{Tr}_{D^*}^{ee}\left(\frac{M^4}{24 m_*^4}-\frac{ M^2}{3 m_*^2}\right),\nonumber\\
&e_1=-\text{Tr}_{D^*}^{em} \left(\frac{M^4}{12 m_*^3}+\frac{M^2}{6 m_*}\right)+\text{Tr}_{D^*}^{mm} \left(-\frac{M^4}{24 m_*^2}
-\frac{13 M^2}{12 }\right)\nonumber\\
&\hspace*{80mm}+\text{Tr}_{D^*}^{ee}\left(-\frac{M^4}{24 m_*^4}+\frac{7 M^2}{12 m_*^2}+\frac{10}{3}\right),\nonumber\\
&e_2=\text{Tr}_{D^*}^{ee}\left(\frac{2m_*^2}{M^2}+\frac{9}{4 }\right)+\frac{m_*^2 \text{Tr}_{D^*}^{mm}}{4 }-\frac{3 m_* \text{Tr}_{D^*}^{em}}{2 },\nonumber\\
&e_3=-\text{Tr}_{D^*}^{em} \left(\frac{13 M^4}{72 m_*^3}-\frac{7 M^2}{6 m_*}\right)+\text{Tr}_{D^*}^{mm} \left(-\frac{13 M^4}{144 m_*^2}
-\frac{17 M^2}{12 }\right)\nonumber\\
&\hspace*{80mm}+\text{Tr}_{D^*}^{ee}\left(-\frac{13 
M^4}{144 m_*^4}+\frac{67 M^2}{36 m_*^2}+\frac{26}{3}\right)\nonumber.
\end{align}
We note that both $C_4[D^*DD^*]$ and $C_4[D^*DD^*]$ contain nonvanishing divergent pieces (proportional to $\lambda$ --- see Appendix~\ref{app:hel0} for the 
definition of $\lambda$) which do not cancel each other in the total helicity-2 amplitude and which, therefore, call for a contact term for renormalisation.

 
\providecommand{\href}[2]{#2}\begingroup\raggedright\endgroup

\end{document}